\documentclass{svjour3}                     
\smartqed  
\usepackage{graphicx}
\usepackage{amsmath,amssymb}

\newcommand{\chiral}{SU(3)$_R\, \times \, $SU(3)$_L$\,}

\journalname{Few-Body Systems (FB20)}
\begin{document}

\title{
Antikaon-nucleon dynamics and its application to few-body systems 
\thanks{Presented at the 20th International IUPAP Conference on Few-Body Problems in Physics, 20 - 25 August, 2012, Fukuoka, Japan. }
}
\subtitle{
}


\author{Tetsuo Hyodo
}


\institute{Tetsuo Hyodo \at
              Department of Physics, Tokyo Institute of Technology,
Tokyo 152-8551, Japan \\
              \email{hyodo@th.phys.titech.ac.jp}           
}

\date{Received: date / Accepted: date}

\maketitle

\begin{abstract}

We overview the recent progress in the investigation of the antikaon dynamics with single nucleon and with few-body nuclei, based on chiral SU(3) symmetry. We first show how the $\Lambda(1405)$ resonance emerges from the coupled-channel $\bar{K}N$-$\pi\Sigma$ interaction. Next, we present the construction of an up-to-date $\bar{K}N$ interaction in response to the recent measurement of kaonic hydrogen. Finally, we discuss the possibility of the antikaon bound states in nuclei from the perspective of few-body physics. 

\end{abstract}

\section{Introduction}
\label{intro}

Antikaon ($\bar{K}$ meson) plays a peculiar role in hadron and nuclear physics. On one hand, $\bar{K}$ is regarded as a pseudo Nambu-Goldstone boson associated with spontaneous breaking of chiral \chiral symmetry. On the other hand, $\bar{K}$ is moderately massive due to the strange quark mass. As a consequence of this dual (chiral/massive) nature, plenty of interesting phenomena have been discussed in the $\bar{K}N$ and $\bar{K}$-nuclear systems~\cite{Hyodo:2011ur}. At the same time, the $\bar{K}N$ interaction is a suitable testing ground for investigating the interplay between spontaneous and explicit chiral symmetry breaking in low-energy QCD. Among others, we here address the following issues:
\begin{itemize}
\item What is the structure of the $\Lambda(1405)$ resonance?
\item How can we construct a realistic $\bar{K}N$-$\pi\Sigma$ interaction?
\item Is $\bar{K}$ bound in few-body nuclei?
\end{itemize}
Through these questions, we introduce recent developments of antikaon physics from theoretical as well as phenomenological viewpoint.

\section{Nature of the $\Lambda(1405)$ resonance}
\label{L1405}

When $\bar{K}$ is combined with one nucleon, it is mandatory to consider the negative parity excited baryon in the isospin $I=0$ channel, the $\Lambda(1405)$ resonance. This resonance locates slightly below the $\bar{K}N$ threshold, and its structure has been an issue in hadron physics for a long time. While the conventional quark model reproduces systematics of the excited baryons, $\Lambda(1405)$ is a well-known exceptional state which largely deviates from the theoretical prediction~\cite{Isgur:1978xj}. In contrast, $\bar{K}$ interacts with the nucleon by strong attraction in the coupled-channel vector-meson-exchange model~\cite{Dalitz:1967fp}, so the $\Lambda(1405)$ resonance can be well described as a quasi-bound state of the $\bar{K}N$ system embedded in the $\pi\Sigma$ continuum.

\begin{figure*}
    \includegraphics[width=0.75\textwidth,bb=-100 0 550 100]{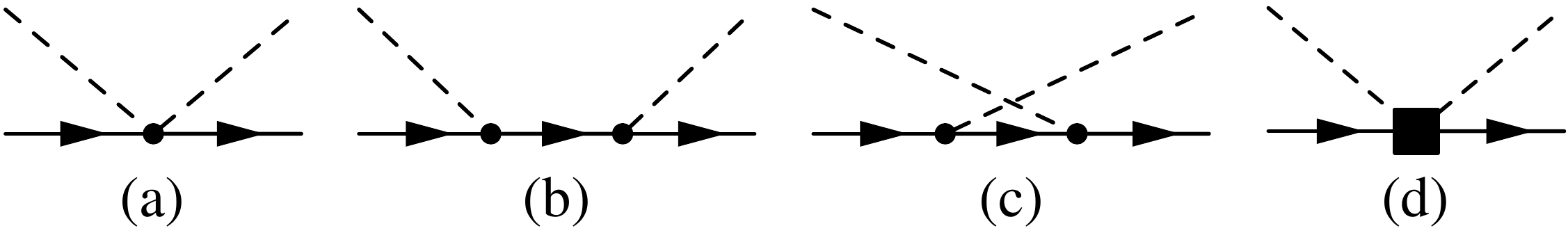}
\caption{Feynman diagrams for the meson-baryon interactions in chiral perturbation theory. (a) Weinberg-Tomozawa interaction, (b) s-channel Born term, (c) u-channel Born term, (d) NLO interaction. The dots represent the $\mathcal{O}(p)$ vertices while the square denotes the $\mathcal{O}(p^2)$ vertex.}
\label{fig:interaction}
\end{figure*}

This latter picture is reformulated from modern viewpoint as the chiral coupled-channel approach~\cite{Kaiser:1995eg,Oset:1998it,Oller:2000fj,Lutz:2001yb,Hyodo:2011ur}, in which the interaction derived from chiral perturbation theory is plugged into the scattering equation. This is in essence similar to the chiral EFT approach for nuclear force~\cite{Machleidt:2011zz}. In the $s$-wave meson-baryon scattering, the interaction kernel $V(W)$ is a function of the total energy $W$, which is put in the scattering equation to obtain the scattering amplitude $T(W)$ as 
\begin{align}
   T(W)
   =&
   \frac{1}{V(W)^{-1}-G(W;a)} ,
   \label{eq:BSE}
\end{align}
where $G(W;a)$ represents the loop integral whose ultraviolet divergence is removed by the dimensional regularization and the finite part is specified by the subtraction constant $a$. The interaction kernel as well as the scattering amplitude are given by matrices in meson-baryon channel space, and the off-diagonal components of $V$ cause the transition between different channels. The interaction kernel can be sorted out by the counting rule in chiral perturbation theory:
\begin{align}
   V
   =&
   V_{\text{TW}}+V_{\text{Born}}
   +V_{\text{NLO}}
   +\dots ,
   \label{eq:interaction}
\end{align}
where $V_{\text{TW}}$ and $V_{\text{Born}}$ are $\mathcal{O}(p)$ terms and next-to-leading order (NLO) interaction $V_{\text{NLO}}$ is counted as $\mathcal{O}(p^{2})$. Diagrammatic expression of each term is shown in Fig.~\ref{fig:interaction}. If the attractive interaction is strong enough, a pole singularity can be generated in the scattering amplitude in Eq.~\eqref{eq:BSE} which is interpreted as a baryon resonance. The most important piece in the leading order $\mathcal{O}(p)$ term is the Tomozawa-Weinberg interaction $V_{\text{TW}}$~\cite{Tomozawa:1966jm,Weinberg:1966kf}. This term shares common features with the vector-meson-exchange potential~\cite{Dalitz:1967fp} through the flavor SU(3) symmetry. It has been shown that $\Lambda(1405)$ and $S=-1$ meson-baryon scattering are well described in this framework~\cite{Kaiser:1995eg,Oset:1998it,Oller:2000fj,Lutz:2001yb,Hyodo:2011ur}.

\begin{figure*}
  \includegraphics[width=0.5\textwidth,bb=-125 0 350 425]{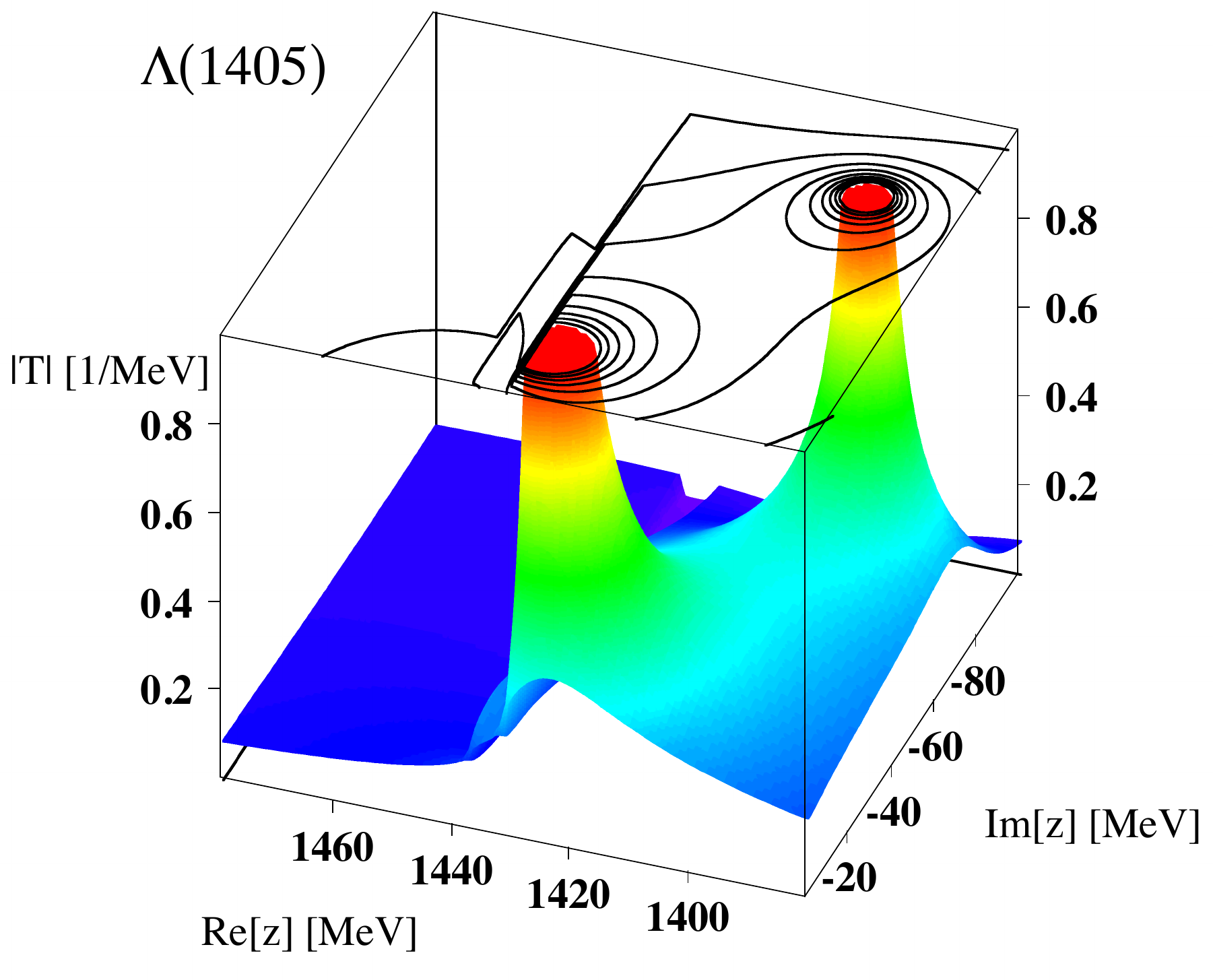}
\caption{Absolute value of the analytic continuation of the scattering amplitude $|T(z)|$ of the $\bar{K}N$ elastic channel in the second Riemann sheet of the complex energy $z$ plane.}
\label{fig:pole1405}
\end{figure*}

An interesting point to be noted here is that $\Lambda(1405)$ is accompanied by two poles~\cite{Jido:2003cb} as shown in Fig.~\ref{fig:pole1405}. The origin of these poles can be traced back to the Tomozawa-Weinberg term for $\bar{K}N$($i=1$)-$\pi\Sigma$($i=2$) systems
\begin{align}
   V_{\text{TW}}
   =&
   -\begin{pmatrix}
   3 & -\sqrt{\frac{3}{2}} \\
   -\sqrt{\frac{3}{2}} & 4
   \end{pmatrix}
   \frac{\omega_{i}+\omega_{j}}{4f^{2}}
   \label{eq:VWT}
\end{align}
where $\omega_{i}$ represents the meson energy in channel $i$ and $f$ is the meson decay constant. We find that the diagonal interactions in both $\bar{K}N$ and $\pi\Sigma$ channels are attractive. Because the Tomozawa-Weinberg interaction was originally derived by the chiral low energy theorem, the interaction strength in Eq.~\eqref{eq:VWT} is model independent as far as we respect chiral symmetry for low energy hadron scattering. Solving the scattering equation~\eqref{eq:BSE} without the off-diagonal components, we obtain a bound state in the $\bar{K}N$ channel and a resonance state in the $\pi\Sigma$ channel~\cite{Hyodo:2007jq}. These are the origins of two poles of $\Lambda(1405)$. In this way, $\Lambda(1405)$ is understood as the $\bar{K}N$ quasi-bound state in the resonating $\pi\Sigma$ continuum. 

An important question is whether the structure of $\Lambda(1405)$ is a two-hadron molecule state or a genuine quark state. This point has been studied from the viewpoint of the natural renormalization scheme~\cite{Hyodo:2008xr}, response to the variation of number of colors $N_{c}$~\cite{Hyodo:2007np,Roca:2008kr}, electromagnetic radii and form factors~\cite{Sekihara:2008qk,Sekihara:2010uz}, and the compositeness and spatial size~\cite{Hyodo:2011qc,Sekihara:2012fu}. All these analyses indicate that $\Lambda(1405)$ is dominated by the meson-baryon molecule component. The study of the $\bar{K}$-nuclear systems in section 4 is also backed by this finding.

\section{Realistic $\bar{K}N$-$\pi\Sigma$ interaction}
\label{IHW}

For the application to the $\bar{K}$ bound states in nuclei, it is indispensable to constrain the $\bar{K}N$ interaction by experimental data quantitatively. In fact, the discrepancy of theoretical predictions of the $\bar{K}NN$ state (see section~\ref{fewbody}) is partly attributed to the uncertainty in the subthreshold extrapolation of the $\bar{K}N$ interaction. To this end, let us classify available experimental data in terms of the relevant energy region with respect to the $\bar{K}N$ threshold energy $W_{\bar{K}N}$ as
\begin{itemize}

\item[1)] total cross sections of $K^{-}p$ scattering ($W>W_{\bar{K}N}$),

\item[2)] threshold branching ratios $\gamma, R_{n}$ and $R_{c}$ ($W=W_{\bar{K}N}$),

\item[3)]  energy shift $\Delta E$ and width $\Gamma$ of kaonic hydrogen ($W=W_{\bar{K}N}$),

\item[4)] $\pi\Sigma$ mass spectrum ($W<W_{\bar{K}N}$).

\end{itemize}
The cross section data 1) is obtained in old bubble chamber experiments for elastic ($\to K^{-}p$) and inelastic ($\to \bar{K}^{0}n, \pi^{+}\Sigma^{-}, \pi^{-}\Sigma^{+}, \pi^{0}\Sigma^{0}, \pi^{0}\Lambda$) channels. Although it constrains the energy dependence of the $\bar{K}N$ interaction above the threshold, the error bars are relatively large and there are some discrepancies among different measurements. The threshold branching ratios 2) are accurately determined from the $K^{-}$ capture by hydrogen, but the data constrains the amplitude only at the threshold $W=W_{\bar{K}N}$. It has been shown in a comprehensive analysis~\cite{Borasoy:2006sr} that the combination of 1) and 2) is not sufficient to predict the $K^{-}p$ scattering length, due to uncertainties of data. It is therefore needed to include other constraints. The energy level of the kaonic hydrogen 1s atomic state 3) is related with the complex $K^{-}p$ scattering length through the improved Deser-Trueman formula. Recently, new high-precision measurement of kaonic hydrogen has been performed by SIDDHARTA~\cite{Bazzi:2011zj}. In view of the accuracy of the SIDDHARTA result, it can be an important constraint on the $\bar{K}N$ interaction.

Persistent experimental efforts have been devoted to the extraction of the $\pi\Sigma$ spectrum 4) where $\Lambda(1405)$ is observed as a peak structure~\cite{Niiyama:2008rt,Moriya:2009mx,Agakishiev:2012xk}. The $\pi\Sigma$ spectrum is also important for the subthreshold $\bar{K}N$ interaction through the coupled-channel effect. It is however not very suitable to include the $\pi\Sigma$ spectrum in the fitting analysis of the meson-baryon amplitude because of the following reasons. First, since the $\pi\Sigma$ elastic scattering is experimentally impossible, the $\pi\Sigma$ spectrum is only accessible via production processes. The spectrum therefore depends on the production mechanism at initial stage as well as the final state interaction with other produced particles. Second, the experimental spectrum contains background contributions which are not completely under control. Third, the magnitude (absolute value) of the spectrum is usually in arbitrary unit. Even if the cross section is measured, it again depends on the model-dependent production mechanism. Hence, there is no ``model-independent'' method to relate the experimentally observed spectrum to the two-body scattering amplitude. These facts hinder a fair analysis of the experimental data. Nevertheless, one should check that the $\pi\Sigma$ spectrum obtained in the analysis is not very much far from the results of various experiments, in order to avoid unphysical solutions. Instead of the $\pi\Sigma$ spectrum, we note that the
\begin{itemize}

\item[5)] $\pi\Sigma$ scattering length ($W=W_{\pi\Sigma}< W_{\bar{K}N}$)

\end{itemize}
can be an alternative constraint to the meson-baryon scattering amplitude. The scattering length is a normalized number at fixed energy ($W=W_{\pi\Sigma}$), and is directly related to the two-body scattering amplitude. It is shown in Ref.~\cite{Ikeda:2011dx} that the subthreshold extrapolation of the $\bar{K}N$ amplitude as well as the structure of $\Lambda(1405)$ are closely related to the value of the $\pi\Sigma$ scattering length. At this moment, there is no experimental information, but determination of the $\pi\Sigma$ scattering length may be possible in the weak decays of $\Lambda_{c}$~\cite{Hyodo:2011js} and also by lattice QCD~\cite{Torok:2009dg,Ikeda:2011qm}.

Based on the above discussion, we perform a systematic $\chi^{2}$ analysis of the meson-baryon scattering amplitude in chiral coupled-channel approach, including experimental data of 1), 2), and the new SIDDHARTA result of 3)~\cite{Ikeda:2011pi,Ikeda:2012au}. The $\pi\Sigma$ spectrum 4) is not used in the fitting procedure, but the spectrum is compared with experimental data in Ref.~\cite{Ikeda:2012au}. To investigate the significance of each term, we systematically include the interaction terms in Eq.~\eqref{eq:interaction} as $V=V_{\text{TW}}$, $V=V_{\text{TW}}+V_{\text{Born}}$, and $V=V_{\text{TW}}+V_{\text{Born}}+V_{\text{NLO}}$. We use physical values for hadron masses and meson decay constants and the meson-baryon axial vector coupling contains $D$ and $F$ are determined by semileptonic hyperon decay. Then, the leading order interactions $V_{\text{TW}}$ and $V_{\text{Born}}$ are completely fixed, while the NLO interaction $V_{\text{NLO}}$ contains seven low-energy constants which are not determined by symmetry argument. In the loop function $G$, there are six subtraction constants $a_{i}$ for each meson-baryon channel $i$. Thus, the number of free parameters is six (thirteen) in the model with (without) the NLO interaction.

\begin{figure}[tb]
\begin{minipage}[t]{6.5cm}
\includegraphics[width=6.5cm,bb=0 0 1200 750]{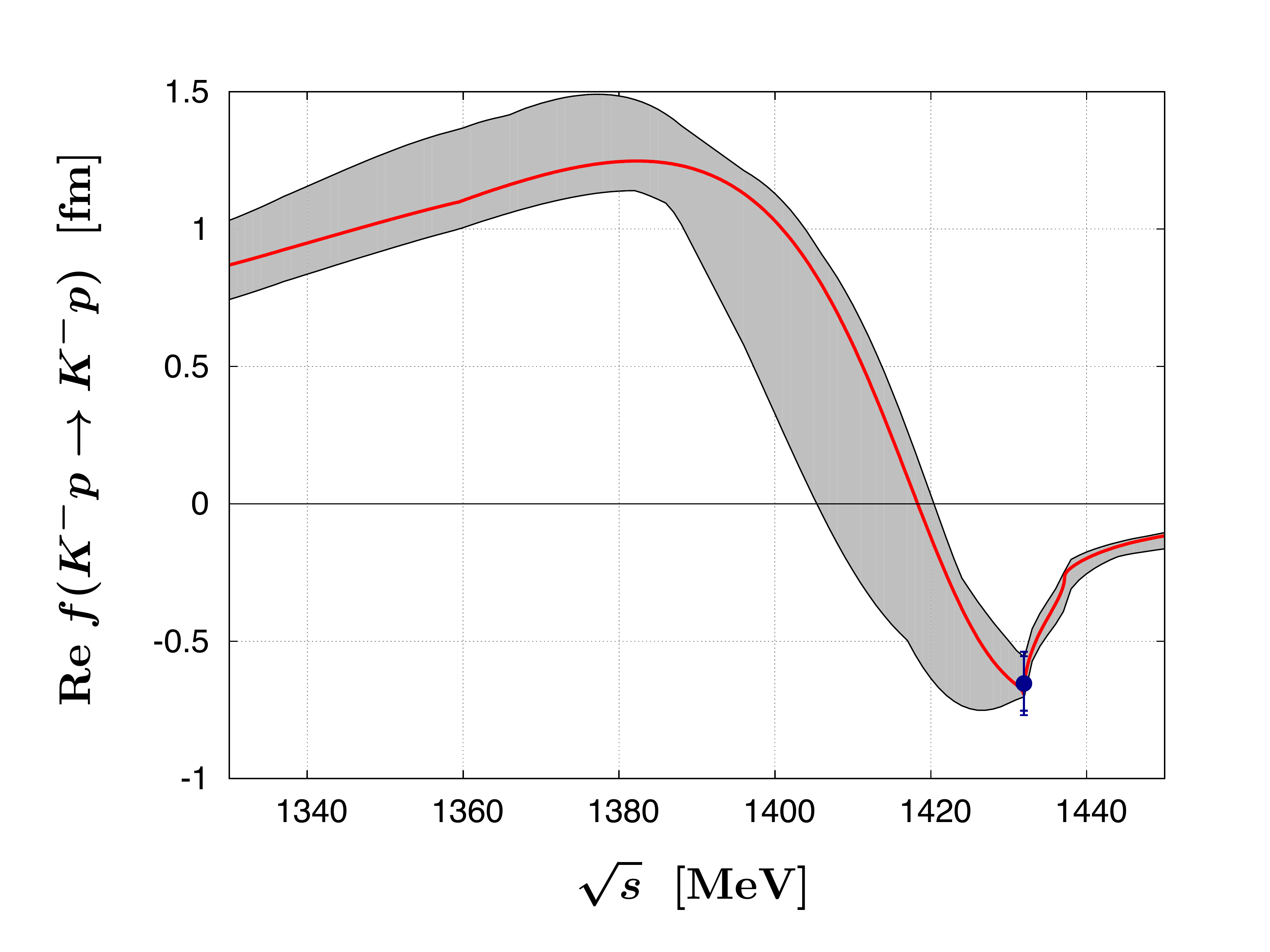}
\end{minipage}
\hspace{-1cm}
\begin{minipage}[t]{6.5cm}
\includegraphics[width=6.5cm,bb=0 0 1200 750]{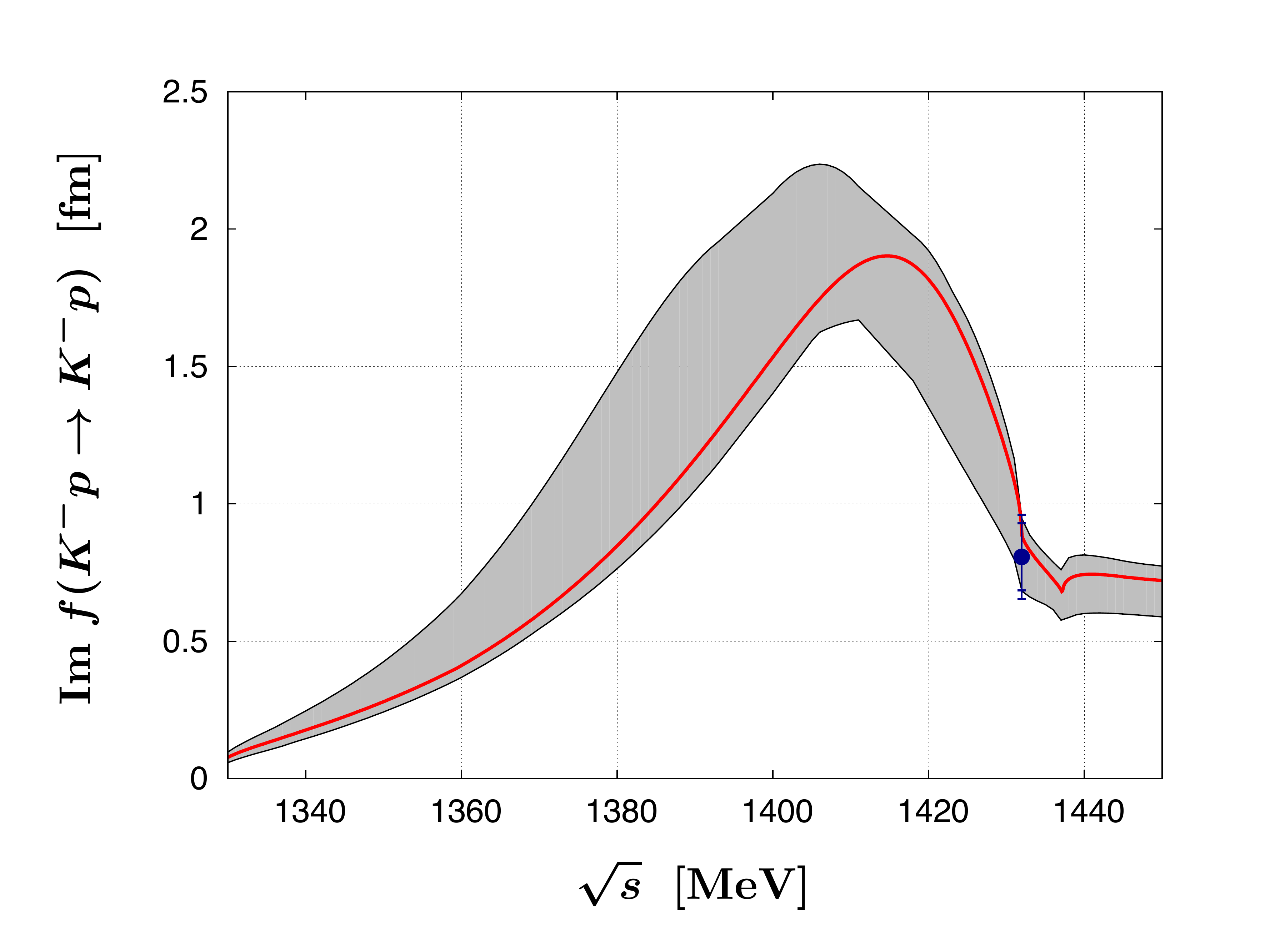}
\end{minipage}
\caption{Real part (left) and imaginary part (right) of the $K^-p \rightarrow K^-p$ forward scattering amplitude extrapolated to the subthreshold region. The empirical real and imaginary parts of the $K^-p$ scattering length deduced from the recent kaonic hydrogen measurement (SIDDHARTA~\cite{Bazzi:2011zj}) are indicated by the dots including statistical and systematic errors. The shaded bands represent theoretical uncertainties.} 
\label{fig:amplitude}
\end{figure}

By performing the $\chi^{2}$ analysis, we find that the models with $V_{\text{TW}}$ and $V_{\text{TW}}+V_{\text{Born}}$ provide reasonable results ($\chi^{2}/\text{d.o.f}=1.12$ and 1.15, respectively), while best-fit requires the contribution from the NLO terms ($\chi^{2}/\text{d.o.f}=0.96$). In this way, we obtain a consistent description of the cross section data together with the $K^{-}p$ scattering length. We also perform the error analysis by introducing small variations of the subtraction constants. In Fig.~\ref{fig:amplitude}, we show the uncertainty region of the subthreshold extrapolation of the $K^{-}p$ elastic scattering amplitude, constrained by the SIDDHARTA result and $K^{-}p\to \pi^{0}\Lambda$ cross sections. In contrast to the previous studies~\cite{Borasoy:2004kk,Borasoy:2005ie}, the uncertainty is significantly reduced and the behavior of the amplitude in the subthreshold energy region is better controlled. This is also important for the study of nonmesonic $\bar{K}$ absorptions in nuclei. In this way, we find that the inclusion of SIDDHARTA result establishes the consistency of the scattering length with the cross section data, and reduces the uncertainty in the subthreshold extrapolation of the $\bar{K}N$ interaction.

%
%

The high-precision data and the systematic analysis also reveal the origin of the remaining uncertainties. For instance, calculating the meson-baryon scattering in charge $Q=-1$ sector with the same model parameters, we find that the scattering length of $K^{-}n$ suffers from relatively large uncertainty. This result indicates that the present experimental database is not sufficient to control the isospin $I=1$ component of the meson-baryon amplitude. The $I=1$ information may be supplied by the measurement of kaonic deuterium, from which the $K^{-}n$ scattering length can be extracted.


\section{Application to few-body systems}
\label{fewbody}

Because the $\Lambda(1405)$ resonance is interpreted as a quasi-bound state of the $\bar{K}N$ system, $\bar{K}$ is considered to form a quasi-bound state also in nuclei~\cite{PL7.288,Akaishi:2002bg}. This can be understood by simple comparison of the $NN$ and $\bar{K}N$ two-body interactions. It is known that two-body forces are attractive in both cases, and there is an $s$-wave bound (quasi-bound) state of the $NN$ ($\bar{K}N$) system in the $I=0$ channel. Thus, we may expect that $\bar{K}$-nuclear systems exhibit rich phenomena in many-body systems. It is interesting to note that the binding energy of the $\Lambda(1405)$ ($\sim$15-30 MeV) is much larger than that of the deuteron ($\sim 2$ MeV). At the same time, one should keep in mind the decay processes; $\bar{K}N$ can decay into $\pi\Sigma$ and $\pi\Lambda$ in free space, and $\bar{K}$ can be absorbed by two or more nucleons in many-body systems.

The simplest $\bar{K}NN$ system has been studied by several groups using different treatments of $\bar{K}N$ interactions and few-body techniques~\cite{Shevchenko:2006xy,Shevchenko:2007zz,Ikeda:2007nz,Yamazaki:2007cs,Dote:2008in,Dote:2008hw,Ikeda:2010tk,Barnea:2012qa}. Because of the ratio of the isospin components of the $\bar{K}N$ pair, spin $J=0$ system (which roughly corresponds to ``$K^{-}pp$'') is intensively studied, while spin $J=1$ state (``$K^{-}d$'') is also discussed~\cite{PL7.288}. In Table~\ref{tab:KbarNN}, we summarize the present status of the theoretical studies of the $\bar{K}NN$ system with total spin $J=0$. In all cases, $\bar{K}NN$ system supports a quasi-bound state, while the results of the binding energy and width are not quantitatively converging. Note however that all the calculations, except for Ref.~\cite{Barnea:2012qa}, have been performed before the SIDDHARTA experiment, so the $K^{-}p$ threshold constraint has not been taken into account. In future investigations, the inclusion of the SIDDHARTA constraint for the $\bar{K}N$ interaction will reduce the uncertainty of theoretical predictions of the $\bar{K}NN$ system.

\begin{table}[bp]
\caption{Theoretical results of the $\bar{K}NN$ system with total spin $J=0$ in the chronological order (from left to right). The binding energy $B_{\bar{K}NN}$ is measured from the $\bar{K}NN$ threshold, and the three-body decay width $\Gamma_{\pi YN}$ does not include the absorption width to $YN$ states. In ``Method'' row, Fadd. (Var.) stands for the Faddeev coupled-channel calculation (variational calculation with effective single-channel potential).
The ``$\bar{K}N$ interaction'' row indicates the energy-(in)dependence of the $\bar{K}N$ interaction. Ref.~\cite{Ikeda:2010tk} found additional pole at lower energy with huge width.} 
\label{tab:KbarNN}       
\begin{tabular}{lllllll}
\hline\noalign{\smallskip}
Reference & \cite{Shevchenko:2006xy,Shevchenko:2007zz} & \cite{Ikeda:2007nz} & \cite{Yamazaki:2007cs} & \cite{Dote:2008in,Dote:2008hw} & \cite{Ikeda:2010tk} & \cite{Barnea:2012qa} \\
\noalign{\smallskip}\hline\noalign{\smallskip}
Method & Fadd. & Fadd. & Var. & Var. & Fadd. & Var. \\
$\bar{K}N$ interaction & E-indep. & E-indep. & E-indep. & E-dep. & E-dep. & E-dep. \\
$B_{\bar{K}NN}$ [MeV] & 50-70 & 60-95 & 48 & 17-23 & 9-16 & 15.7 \\
$\Gamma_{\pi YN}$ [MeV] & $\sim 100$ & 45-80 & 61 & 40-70 & 34-46 & 41.2 \\
\noalign{\smallskip}\hline
\end{tabular}
\end{table}

Spin $J=1$ $\bar{K}NN$ state has been recently studied in variational calculation with $\Lambda^{*}N$ potential picture~\cite{Uchino:2011jt}, fixed center approximation to Faddeev equation~\cite{Oset:2012gi}, and also three-body variational calculation of the $\bar{K}NN$ system~\cite{Barnea:2012qa}. Note that the former two studies have introduced an approximation in three body calculations; either the model space is reduced by assuming the $\Lambda^{*}$ cluster of the $\bar{K}N$ system, or the explicit $NN$ dynamics is not solved in the calculation. In Refs.~\cite{Uchino:2011jt,Barnea:2012qa}, the lowest energy state of spin $J=1$ channel turns out to be $\Lambda^{*}N$ scattering state and no bound state is found below that. This does not contradict to Ref.~\cite{Oset:2012gi} where a quasi-bound state is found at 9 MeV below the $\bar{K}NN$ threshold, because the variational study searches for the ground state of the system. More detailed quest for the quasi-bound state in this channel is interesting, since the binding energy is expected to be small and the quasi-bound state is closely related to the $\bar{K}N$ threshold information.

Before closing this section, we briefly mention an analogues system in the charm sector, the $DN$ interaction and the $DNN$ quasi-bound state~\cite{Bayar:2012dd}. Using the vector meson exchange interaction~\cite{Mizutani:2006vq}, we can identify the negative parity $\Lambda_{c}(2595)$ resonance as a $DN$ quasi-bound state. In this case, because of the narrow width of $\Lambda_{c}(2595)$ ($\sim 3$ MeV), the width of the three-body quasi-bound state can also be small~\cite{Bayar:2012dd}, which is advantageous in experimental observation.

\section{Conclusion}
\label{conclusion}

We have studied the $\bar{K}N$-$\pi\Sigma$ interaction and its applications to few-body systems. It is found that the strong coupled-channel $\bar{K}N$-$\pi\Sigma$ dynamics generates the $\Lambda(1405)$ resonance as a quasi-bound $\bar{K}N$ state in the strongly correlated $\pi\Sigma$ continuum. We show that the recent measurement of kaonic hydrogen provides a strong constraint on the $\bar{K}N$-$\pi\Sigma$ interaction. Theoretical investigations of the $\bar{K}NN$ system suggest that the antikaon should be bound in few-body nuclear systems, although the quantitative results have been scattered due to experimental uncertainties. Thus, we are now in a position to study various interesting phenomena of $\bar{K}$ quasi-bound states in few-nucleon systems, using the realistic $\bar{K}N$-$\pi\Sigma$ interaction constrained by the kaonic hydrogen measurement.

\begin{acknowledgements}
The author is grateful to Yoichi Ikeda, Daisuke Jido and Wolfram Weise for many discussions and collaborations.
This work is partly supported by the Grant-in-Aid for Scientific Research from MEXT and JSPS (No. 24105702 and No. 24740152)
and the Global Center of Excellence Program by MEXT, Japan, through the Nanoscience and Quantum Physics Project of the Tokyo Institute of Technology.

\end{acknowledgements}


\begin{thebibliography}{10}
\providecommand{\url}[1]{{#1}}
\providecommand{\urlprefix}{URL }
\expandafter\ifx\csname urlstyle\endcsname\relax
  \providecommand{\doi}[1]{DOI~\discretionary{}{}{}#1}\else
  \providecommand{\doi}{DOI~\discretionary{}{}{}\begingroup
  \urlstyle{rm}\Url}\fi

\bibitem{Hyodo:2011ur}
Hyodo, T., Jido, D.: {The nature of the $\Lambda(1405)$ resonance in chiral
  dynamics}.
\newblock Prog. Part. Nucl. Phys. 67, 55--98 (2012)

\bibitem{Isgur:1978xj}
Isgur, N., Karl, G.: {$P$-wave baryons in the quark model}.
\newblock Phys. Rev. D18, 4187 (1978)

\bibitem{Dalitz:1967fp}
Dalitz, R.H., Wong, T.C., Rajasekaran, G.: {Model calculation for $Y^*_0
  (1405)$ resonance state}.
\newblock Phys. Rev. 153, 1617--1623 (1967)

\bibitem{Kaiser:1995eg}
Kaiser, N., Siegel, P.B., Weise, W.: Chiral dynamics and the low-energy
  kaon--nucleon interaction.
\newblock Nucl. Phys. A594, 325--345 (1995)

\bibitem{Oset:1998it}
Oset, E., Ramos, A.: {Non perturbative chiral approach to s-wave $\bar{K} N$
  interactions}.
\newblock Nucl. Phys. A635, 99--120 (1998)

\bibitem{Oller:2000fj}
Oller, J.A., Meissner, U.G.: Chiral dynamics in the presence of bound states:
  kaon--nucleon interactions revisited.
\newblock Phys. Lett. B500, 263--272 (2001)

\bibitem{Lutz:2001yb}
Lutz, M.F.M., Kolomeitsev, E.E.: {Relativistic chiral SU(3) symmetry, large
  $N_c$ sum rules and meson baryon scattering}.
\newblock Nucl. Phys. A700, 193--308 (2002)

\bibitem{Machleidt:2011zz}
Machleidt, R., Entem, D.: {Chiral effective field theory and nuclear forces}.
\newblock Phys. Rept. 503, 1--75 (2011)

\bibitem{Tomozawa:1966jm}
Tomozawa, Y.: Axial vector coupling renormalization and the meson baryon
  scattering lengths.
\newblock Nuovo Cim. 46A, 707--717 (1966)

\bibitem{Weinberg:1966kf}
Weinberg, S.: Pion scattering lengths.
\newblock Phys. Rev. Lett. 17, 616--621 (1966)

\bibitem{Jido:2003cb}
Jido, D., Oller, J.A., Oset, E., Ramos, A., Meissner, U.G.: {Chiral dynamics of
  the two $\Lambda(1405)$ states}.
\newblock Nucl. Phys. A725, 181--200 (2003)

\bibitem{Hyodo:2007jq}
Hyodo, T., Weise, W.: {Effective $\bar{K} N$ interaction based on chiral SU(3)
  dynamics}.
\newblock Phys. Rev. C77, 035204 (2008)

\bibitem{Hyodo:2008xr}
Hyodo, T., Jido, D., Hosaka, A.: {Origin of resonances in the chiral unitary
  approach}.
\newblock Phys. Rev. C78, 025203 (2008)

\bibitem{Hyodo:2007np}
Hyodo, T., Jido, D., Roca, L.: {Structure of the $\Lambda(1405)$ baryon
  resonance from its large $N_c$ behavior}.
\newblock Phys. Rev. D77, 056010 (2008)

\bibitem{Roca:2008kr}
Roca, L., Hyodo, T., Jido, D.: {On the nature of the $\Lambda(1405)$ and
  $\Lambda(1670)$ from their $N_c$ behavior in chiral dynamics}.
\newblock Nucl. Phys. A809, 65--87 (2008)

\bibitem{Sekihara:2008qk}
Sekihara, T., Hyodo, T., Jido, D.: {Electromagnetic mean squared radii of
  $\Lambda(1405)$ in chiral dynamics}.
\newblock Phys. Lett. B669, 133--138 (2008)

\bibitem{Sekihara:2010uz}
Sekihara, T., Hyodo, T., Jido, D.: {Internal structure of resonant
  $\Lambda(1405)$ state in chiral dynamics}.
\newblock Phys. Rev. C83, 055202 (2011)

\bibitem{Hyodo:2011qc}
Hyodo, T., Jido, D., Hosaka, A.: {Compositeness of dynamically generated states
  in a chiral unitary approach}.
\newblock Phys. Rev. C85, 015201 (2012)

\bibitem{Sekihara:2012fu}
Sekihara, T., Hyodo, T.: {Size measurement of dynamically generated resonances
  with finite boxes}. arXiv:1209.0577 [nucl-th]

\bibitem{Borasoy:2006sr}
Borasoy, B., Meissner, U.G., Nissler, R.: {$K^- p$ scattering length from
  scattering experiments}.
\newblock Phys. Rev. C74, 055201 (2006)

\bibitem{Bazzi:2011zj}
Bazzi, M., et~al.: {A New Measurement of Kaonic Hydrogen X-rays}.
\newblock Phys. Lett. B704, 113--117 (2011)

\bibitem{Niiyama:2008rt}
Niiyama, M., et~al.: {Photoproduction of $\Lambda(1405)$ and $\Sigma^{0}(1385)$
  on the proton at $E_\gamma = 1.5$--2.4 GeV}.
\newblock Phys. Rev. C78, 035202 (2008)

\bibitem{Moriya:2009mx}
Moriya, K., Schumacher, R.: {Properties of the $\Lambda(1405)$ Measured at
  CLAS}.
\newblock Nucl. Phys. A835, 325--328 (2010)

\bibitem{Agakishiev:2012xk}
Agakishiev, G., et~al.:
  {Baryonic resonances close to the $\bar{K}N$ threshold: the case of
  $\Lambda(1405)$ in pp collisions}. arXiv:1208.0205 [nucl-ex]

\bibitem{Ikeda:2011dx}
Ikeda, Y., Hyodo, T., Jido, D., Kamano, H., Sato, T., Yazaki, K.: {Structure of
  $\Lambda(1405)$ and threshold behavior of $\pi\Sigma$ scattering}.
\newblock Prog. Theor. Phys. 124, 1205--1224 (2011)

\bibitem{Hyodo:2011js}
Hyodo, T., Oka, M.: {Determination of the $\pi \Sigma$ scattering lengths from
  the weak decays of $\Lambda_c$}.
\newblock Phys. Rev. C84, 035201 (2011)

\bibitem{Torok:2009dg}
Torok, A., Beane, S.R., Detmold, W., Luu, T.C., Orginos, K., et~al.:
  {Meson-Baryon Scattering Lengths from Mixed-Action Lattice QCD}.
\newblock Phys. Rev. D81, 074506 (2010)

\bibitem{Ikeda:2011qm}
Ikeda, Y.: {S-wave meson-baryon potentials with strangeness from Lattice QCD}.
\newblock PoS LATTICE2011, 159 (2011)

\bibitem{Ikeda:2011pi}
Ikeda, Y., Hyodo, T., Weise, W.: {Improved constraints on chiral SU(3) dynamics
  from kaonic hydrogen}.
\newblock Phys. Lett. B706, 63--67 (2011)

\bibitem{Ikeda:2012au}
Ikeda, Y., Hyodo, T., Weise, W.: {Chiral SU(3) theory of antikaon--nucleon
  interactions with improved threshold constraints}.
\newblock Nucl. Phys. A881, 98--114 (2012)

\bibitem{Borasoy:2004kk}
Borasoy, B., Nissler, R., Weise, W.: Kaonic hydrogen and k- p scattering.
\newblock Phys. Rev. Lett. 94, 213401 (2005)

\bibitem{Borasoy:2005ie}
Borasoy, B., Nissler, R., Weise, W.: Chiral dynamics of kaon nucleon
  interactions, revisited.
\newblock Eur. Phys. J. A25, 79--96 (2005)

\bibitem{PL7.288}
Nogami, Y.: {Possible existence of $\bar{K}NN$ bound states}.
\newblock Phys. Lett. 7, 288--289 (1963)

\bibitem{Akaishi:2002bg}
Akaishi, Y., Yamazaki, T.: {Nuclear $K$ bound states in light nuclei}.
\newblock Phys. Rev. C65, 044005 (2002)

\bibitem{Shevchenko:2006xy}
Shevchenko, N.V., Gal, A., Mares, J.: {Faddeev calculation of a $K^- p p$
  quasi-bound state}.
\newblock Phys. Rev. Lett. 98, 082,301 (2007)

\bibitem{Shevchenko:2007zz}
Shevchenko, N.V., Gal, A., Mares, J., Revai, J.: {$\bar{K}NN$ quasi-bound state
  and the $\bar{K}N$ interaction: coupled-channel Faddeev calculations of the
  $\bar{K}NN - \pi \Sigma N$ system}.
\newblock Phys. Rev. C76, 044004 (2007)

\bibitem{Ikeda:2007nz}
Ikeda, Y., Sato, T.: {Strange dibaryon resonance in the $\bar{K} N N$ - $\pi
  \Sigma N$ system}.
\newblock Phys. Rev. C76, 035203 (2007)

\bibitem{Yamazaki:2007cs}
Yamazaki, T., Akaishi, Y.: {The basic $\bar{K}$ nuclear cluster $K^- pp$ and
  its enhanced formation in the $p + p \to K^+ + X$ reaction}.
\newblock Phys. Rev. C76, 045201 (2007)

\bibitem{Dote:2008in}
Dote, A., Hyodo, T., Weise, W.: {$K^-pp$ system with chiral SU(3) effective
  interaction}.
\newblock Nucl. Phys. A804, 197--206 (2008)

\bibitem{Dote:2008hw}
Dote, A., Hyodo, T., Weise, W.: {Variational calculation of the $ppK^-$ system
  based on chiral SU(3) dynamics}.
\newblock Phys. Rev. C79, 014003 (2009)

\bibitem{Ikeda:2010tk}
Ikeda, Y., Kamano, H., Sato, T.: {Energy dependence of $\bar{K}N$ interactions
  and resonance pole of strange dibaryons}.
\newblock Prog. Theor. Phys. 124, 533--539 (2010)

\bibitem{Barnea:2012qa}
Barnea, N., Gal, A., Liverts, E.: {Realistic calculations of $\bar{K} N N$,
  $\bar{K} N N N$, and $\bar{K} \bar{K} N N$ quasibound states}.
\newblock Phys. Lett. B712, 132--137 (2012)

\bibitem{Uchino:2011jt}
Uchino, T., Hyodo, T., Oka, M.: {The $\Lambda^* N$ interaction and two-body
  bound state based on chiral dynamics}.
\newblock Nucl. Phys. A868-869, 53--81 (2011)

\bibitem{Oset:2012gi}
Oset, E., et~al.: {A new perspective on the Faddeev equations and the
  $\bar{K}NN$ system from chiral dynamics and unitarity in coupled channels}.
\newblock Nucl. Phys. A881, 127--140 (2012)

\bibitem{Bayar:2012dd}
Bayar, M., Xiao, C.W., Hyodo, T., Dote, A., Oka, M., Oset, E.: {A narrow $DNN$
  quasi-bound state}. Phys. Rev. C86, 044004 (2012)

\bibitem{Mizutani:2006vq}
Mizutani, T., Ramos, A.: {$D$ mesons in nuclear matter: A $D N$ coupled-channel
  equations approach}.
\newblock Phys. Rev. C74, 065201 (2006)

\end{thebibliography}

\end{document}